\documentclass[12pt,english]{article}

\usepackage[T1]{fontenc}
\usepackage[latin9]{inputenc}
\usepackage{geometry}
\usepackage{amsmath}
\usepackage{setspace}
\usepackage{amssymb}

\makeatletter
\newcommand{\lyxaddress}[1]{
\par {\raggedright #1
\vspace{1.4em}
\noindent\par}
}

\usepackage{babel}
\makeatother

\begin{document}

\title{Discrete Symmetries and Generalized Fields of Dyons}

\author{P. S. Bisht$^{\text{(1,2)}},$ Tianjun Li$^{\text{(2)}}$, Pushpa$^{\text{(1)}}$
and O. P. S. Negi$^{\text{(1)}}$}

\maketitle
\begin{singlespace}

\lyxaddress{\begin{center}
1. Department of Physics,\\
 Kumaun University,\\
 S. S. J. Campus, \\
Almora- 263601(Uttarakhand), India
\par\end{center}}
\end{singlespace}

\lyxaddress{\begin{center}
2. Institute of Theoretical Physics,\\
 Chinese Academy of Sciences,\\
 P. O. Box 2735,\\
 Beijing 100080, P. R. China 
\par\end{center}}

\lyxaddress{\begin{center}
Email: ps\_bisht123@rediffmail.com, \\
tli@itp.ac.cn, \\
pushpakalauni60@yahoo.co.in, \\
 ops\_negi@yahoo.co.in
\par\end{center}}

\begin{abstract}
We have studied the different symmetric properties of the generalized
Maxwell's - Dirac equation along with their quantum properties. Applying
the parity ($\mathcal{P}$), time reversal\textbf{ (}$\mathcal{T}$\textbf{)},
charge conjugation ($\mathcal{C}$) and their combined effect like
parity time reversal ($\mathcal{PT}$), charge conjugation and parity
($\mathcal{CP}$) and $\mathcal{CP}T$ transformations to varius equations
of generalized fields of dyons, it is shown that the corresponding
dynamical quantities and equations of dyons are invariant under these
discrete symmetries.


\end{abstract}

\section{Introduction}

In developing the standard model for particles, certain types of interactions
and decays are observed to be common and others seem to be forbidden.
The study of interactions has led to a number of conservation laws
which govern them \cite{key-1,key-2,key-3,key-4}. These conservation
laws are in addition to the classical conservation laws such as conservation
of energy, charge, etc., which still apply in the realm of particle
interactions. Specific quantum numbers have been assigned to the different
fundamental particles, and other conservation laws are associated
with those quantum numbers. From another point of view, it would seem
that any localized particle of finite mass should be unstable, since
the decay into several smaller particles provides many more ways to
distribute the energy, and thus would have higher entropy. This idea
is even stated as a principle called the \char`\"{}totalitarian principle\char`\"{}
which might be stated as \char`\"{}every process that is not forbidden
must occur\char`\"{}. From this point of view, any decay process which
is expected but not observed must be prevented from occurring by some
conservation law. This approach has been fruitful in helping to determine
the rules for particle decay. Conservation laws for parity, isospin,
and strangeness have been developed by detailed observation of particle
interactions \cite{key-1,key-2,key-3,key-4}. The implications of
parity symmetry in atoms were first investigated by Wigner \cite{key-5},
who demonstrated that the combination of charge conjugation $(\mathcal{C})$,
parity $(\mathcal{P})$ and time reversal $(\mathcal{T})$ is considered
to be a fundamental symmetry operation - all physical particles and
interactions appear to be invariant under this combination. Called
$\mathcal{CPT}$ invariance, this symmetry plumbs the depths of our
understanding of nature. Many of the researchers already described
the behavior of these symmetries in electromagnetic fields \cite{key-6,key-7,key-8,key-9,key-10,key-11,key-12,key-13,key-14,key-15,key-16,key-17}.
In a series of papers \cite{key-18,key-19,key-20,key-21,key-22},
we have discussed the generalized Dirac - Maxwell (GDM) equations
in presence of electric and magnetic sources in an isotropic (homogeneous)
medium. We have also analyzed the other quantum equations of dyons
in consistent and manifest covariant way \cite{key-18}. This theory
has been shown to remain invariant under the duality transformations
in isotropic homogeneous medium. Quaternion analysis of time dependent
Maxwell\textquoteright{}s equations has been developed \cite{key-19}
in presence of electric and magnetic charges and the solution for
the classical problem of moving charge (electric and magnetic) are
obtained consistently. The time dependent generalized Dirac - Maxwell\textquoteright{}s
(GDM) equations of dyons have also been discussed \cite{key-20} in
chiral and inhomogeneous media and the solutions for the classical
problem are obtained. The quaternion reformulation of generalized
electromagnetic fields of dyons in chiral and inhomogeneous media
has also been analyzed \cite{key-21}. We have also discussed \cite{key-22}
the monochromatic fields of generalized electromagnetic fields of
dyons in slowly changing media in a consistent manner. In this paper,
we have made an attempt to study the different symmetric properties
of generalized Maxwell's Dirac equations in the homogeneous medium.
It is shown that the generalized Maxwell's - Dirac (GDM) equations
are invariant under discrete symmetries like parity $\mathcal{P}$,
time reversal $\mathcal{T}$, charge conjugation $(\mathcal{C})$
, parity - time reversal operation $\mathcal{PT}$ , charge conjugation
- parity operation $\mathcal{CP}$ and combined operation $(\mathcal{CPT})$.
But when we take the compact form of these equations as complex ones
for generalized fields of dyons, we observe that field equation and
equation of motion of dyons are no more invariant under $\mathcal{P}$
symmetry, $\mathcal{T}$ transformations and combined effects of $\mathcal{CP}$
and $\mathcal{PT}$. On the other hand, these equations and other
quantum equations of dyons in homogeneous medium are invariant under
the combined operation $(\mathcal{CPT})$. As such, we may conclude
that $\mathcal{CPT}$ invariance is an exact symmetry for generalized
fields of dyons and reproduces the consistent theory of electric charge
(magnetic monopole) in the absence magnetic monopole (electric charge
) on dyons.

\section{Fields Associated with dyons}

Assuming the existence of magnetic monopoles, Dirac \cite{key-23}
generalized the Maxwell's equations in homogeneous isotropic medium
written in the following manner in SI units $(c=\hbar=1)$ i. e.

\begin{eqnarray}
\overrightarrow{\nabla.} & \overrightarrow{E}= & \frac{\rho_{e}}{\varepsilon};\nonumber \\
\overrightarrow{\nabla.} & \overrightarrow{B} & =\mu\rho_{m};\nonumber \\
\overrightarrow{\nabla}\times\overrightarrow{E}=-\frac{\partial\overrightarrow{B}}{\partial t} & - & \frac{\overrightarrow{j_{m}}}{\varepsilon};\nonumber \\
\overrightarrow{\nabla}\times\overrightarrow{B}=\frac{1}{v^{2}}\frac{\partial\overrightarrow{E}}{\partial t} & +\mu & \overrightarrow{j_{e}};\label{eq:1}\end{eqnarray}
where $\rho_{e}$ and $\rho_{m}$ are respectively the electric and
magnetic charge densities while $\overrightarrow{j_{e}}$ and $\overrightarrow{j_{m}}$
are the corresponding current densities ,$\varepsilon$ and $\mu$
are defined respectively as relative permittivity and permeability
in electric and magnetic fields in the medium, $\vec{E}$ is electric
field, $\vec{B}$ is magnetic field and $v=\frac{1}{\sqrt{\mu\varepsilon}}$
is the velocity of the electromagnetic wave. Differential equation
(\ref{eq:1}) are the generalized field equation of dyon in homogeneous
medium and the corresponding electric and magnetic fields are called
the generalized electromagnetic fields of dyons. These electromagnetic
fields of dyons are expressed in following differential form in homogeneous
medium in terms of two four potentials as \cite{key-19},

\begin{eqnarray}
\overrightarrow{E} & = & -\overrightarrow{\nabla}\phi_{e}-\frac{\partial\overrightarrow{C}}{\partial t}-\overrightarrow{\nabla}\times\overrightarrow{D};\nonumber \\
\overrightarrow{B} & = & -\overrightarrow{\nabla}\phi_{m}-\frac{1}{v^{2}}\frac{\partial\overrightarrow{D}}{\partial t}+\overrightarrow{\nabla}\times\overrightarrow{C};\label{eq:2}\end{eqnarray}
where $\left\{ C^{\mu}\right\} $=$\left\{ \phi_{e},\, v\overrightarrow{C}\right\} $
and $\left\{ D^{\mu}\right\} $=$\left\{ v\phi_{m},\overrightarrow{\, D}\right\} $
are the two four - potentials respectively associated with electric
and magnetic charges. Let us define the complex vector field $\vec{\psi}$
in the following form,

\begin{eqnarray}
\overrightarrow{\psi}= & \overrightarrow{E}-iv & \overrightarrow{B}.\label{eq:3}\end{eqnarray}
Equations (\ref{eq:1},\ref{eq:2}) and (\ref{eq:3}) , thus give
rise to the following relation between generalized field and the components
of the generalized four - potential as;

\begin{eqnarray}
\overrightarrow{\psi}=-\frac{\partial\overrightarrow{V}}{\partial t}-\overrightarrow{\nabla}\phi-iv(\overrightarrow{\nabla} & \times & \overrightarrow{V}).\label{eq:4}\end{eqnarray}
Here $\left\{ V_{\mu}\right\} $ is the generalized four - potential
of dyons in homogeneous medium and defined as ;

\begin{eqnarray}
\left\{ V_{\mu}\right\}  & = & \left\{ \phi,-\overrightarrow{V}\,\right\} ;\label{eq:5}\end{eqnarray}
where

\begin{eqnarray}
\phi=\phi_{e}- & i & v\phi_{m}\label{eq:6}\end{eqnarray}
and

\begin{eqnarray}
\overrightarrow{V}= & \overrightarrow{C}-i & \frac{\overrightarrow{D}}{v}.\label{eq:7}\end{eqnarray}
Using equations (\ref{eq:3},\ref{eq:4},\ref{eq:5},\ref{eq:6}),
Dirac-Maxwell field equation ( \ref{eq:1} ) may be written in terms
of generalized field $\overrightarrow{\psi}$ as;

\begin{eqnarray}
\overrightarrow{\nabla}.\overrightarrow{\psi} & = & \frac{\rho}{\varepsilon};\nonumber \\
\overrightarrow{\nabla} & \times\overrightarrow{\psi}=-iv(\mu\overrightarrow{J}+\frac{1}{v^{2}} & \frac{\partial\vec{\psi}}{\partial t});\label{eq:8}\end{eqnarray}
where $\rho$ and $\vec{J}$ are respectively the generalized charge
and current source densities of dyons in homogeneous medium \cite{key-19}
given by;

\begin{eqnarray}
\rho=\rho_{e} & -i & \frac{\rho_{m}}{v};\nonumber \\
J=j_{e} & -iv & j_{m}.\label{eq:9}\end{eqnarray}
With the use of equations (\ref{eq:8}) and (\ref{eq:9}), we introduce
a new parameter $\overrightarrow{S}$ ( i.e. the field current) as

\begin{eqnarray}
\overrightarrow{S}=\square\overrightarrow{\psi} & = & -\mu\frac{\partial\overrightarrow{J}}{\partial t}-\frac{1}{\varepsilon}\overrightarrow{\nabla}\rho-iv\mu(\overrightarrow{\nabla}\times\overrightarrow{J});\label{eq:10}\end{eqnarray}
where $\square$ is the D'Alembertian operator expressed as

\begin{eqnarray}
\square\overrightarrow{\psi}== & \frac{1}{v^{2}}\frac{\partial^{2}}{\partial t^{2}}-\nabla^{2}= & \frac{1}{v^{2}}\frac{\partial^{2}}{\partial t^{2}}-\frac{\partial^{2}}{\partial x^{2}}-\frac{\partial^{2}}{\partial y^{2}}-\frac{\partial^{2}}{\partial z^{2}},\label{eq:11}\end{eqnarray}
where $v$ is the speed of electromagnetic wave in homogeneous isotropic
medium. In terms of complex potential, the field equation (\ref{eq:1})
is written as 

\begin{eqnarray}
\square\phi & = & v\mu\rho;\nonumber \\
\square\overrightarrow{V} & = & \mu\overrightarrow{J}.\label{eq:12}\end{eqnarray}
Hence, we may write the tensorial form of generalized Maxwell Dirac
equation (\ref{eq:1}) of dyons in isotropic homogeneous medium as,

\begin{eqnarray}
F_{\mu\nu,\nu} & = & j_{\mu}^{e};\nonumber \\
F_{\mu\nu,\nu}^{d} & = & j_{\mu}^{m};\label{eq:13}\end{eqnarray}
where

\begin{eqnarray}
\left\{ j_{\mu}^{e}\right\}  & = & \left\{ v\rho_{e},\,\overrightarrow{j_{e}}\right\} ;\nonumber \\
\left\{ j_{\mu}^{m}\right\}  & = & \left\{ \rho_{m},\, v\overrightarrow{j_{m}}\right\} .\label{eq:14}\end{eqnarray}
 Defining the generalized field tensor of dyons \cite{key-19} as;

\begin{eqnarray}
G_{\mu\nu,\nu} & =F_{\mu\nu}-i & vF_{\mu\nu}^{d};\label{eq:15}\end{eqnarray}
one can directly obtain the following form of generalized field equation
of dyon in homogeneous isotropic medium  as 

\begin{eqnarray}
G_{\mu\nu,\nu} & = & J_{\mu};\label{eq:16}\end{eqnarray}
where \begin{eqnarray}
\left\{ J_{\mu}\right\}  & = & \left\{ \rho,-\vec{J}\right\} .\label{eq:17}\end{eqnarray}
The Lorentz four - force equation of motion for dyon in homogeneous
isotropic medium is written as;

\begin{eqnarray}
f_{\mu}=m_{0}\ddot{x}_{\mu}= & Re & Q^{*}(G_{\mu\nu}u^{\nu});\label{eq:18}\end{eqnarray}
where $'Re'$ denotes the real part , $\{\ddot{x}_{\mu}\}$ is the
four - acceleration and $\left\{ u^{\nu}\right\} $ is the four -
velocity of the particle and $Q$ is the generalized charge of dyon
in homogeneous isotropic medium .

\section{Effect of Parity on Generalized fields of dyons}

Parity is the symmetry of interaction which involves a transformation
that changes the algebraic sign of the coordinate system \cite{key-1}.
Under the parity transformation $\mathcal{P}$;

\begin{eqnarray}
\mathcal{P}:\left(\begin{array}{c}
x\\
y\\
z\end{array}\right) & \mapsto & \left(\begin{array}{c}
-x\\
-y\\
-z\end{array}\right).\label{eq:19}\end{eqnarray}
The parity transformation changes a right - handed coordinate system
into a left - handed one or vice versa. Under the parity transformations
the physical quantities like time $(t)$, mass $(m)$, electric charge
density $(\rho_{e})$, magnetic current density $(\overrightarrow{j_{m}})$,
electric scalar potential $(\phi_{e})$, magnetic vector potential
$(\overrightarrow{D})$, generalized scalar potential $(\phi)$, time
derivative $(\partial_{t})$, D' Alembertian $(\square)$ and magnetic
field $(\overrightarrow{B})$, have even parity as they do not change
\cite{key-16} their sign under spatial inversion (Parity transformation)
showing that they have even parity.

\begin{eqnarray}
\mathcal{P}(t)\mathcal{P}^{-1} & = & t;\nonumber \\
\mathcal{P}(m)\mathcal{P}^{-1} & = & m;\nonumber \\
\mathcal{P}(\rho_{e})\mathcal{P}^{-1} & = & \rho_{e};\nonumber \\
\mathcal{P}(\overrightarrow{j_{m}})\mathcal{P}^{-1} & = & \overrightarrow{j_{m}};\nonumber \\
\mathcal{P}(\phi_{e})\mathcal{P}^{-1} & = & \phi_{e};\nonumber \\
\mathcal{P}(\overrightarrow{D})\mathcal{P}^{-1} & = & \overrightarrow{D};\nonumber \\
\mathcal{P}(\partial_{t})\mathcal{P}^{-1} & = & \partial_{t};\nonumber \\
\mathcal{P}(\square)\mathcal{P}^{-1} & = & \square;\nonumber \\
\mathcal{P}(\overrightarrow{B})\mathcal{P}^{-1} & = & \overrightarrow{B}.\label{eq:20}\end{eqnarray}
On the other hand, the physical like displacement $(x)$, velocity
$(\overrightarrow{u})$, acceleration $(\overrightarrow{a})$, magnetic
charge density $(\rho_{m})$, electric current density $(\overrightarrow{j_{e}})$,
magnetic scalar potential $(\phi_{m})$, differential operator $(\overrightarrow{\nabla})$,
electric vector potential $(\overrightarrow{C})$ and electric field
$(\overrightarrow{E})$ have odd parity as they change \cite{key-16}
their sign under spatial inversion (Parity transformation) i.e.

\begin{eqnarray}
\mathcal{P}(x)\mathcal{P}^{-1} & = & -x;\nonumber \\
\mathcal{P}(\overrightarrow{u})\mathcal{P}^{-1} & = & -\overrightarrow{u};\nonumber \\
\mathcal{P}(\overrightarrow{a})\mathcal{P}^{-1} & = & -\overrightarrow{a};\nonumber \\
\mathcal{P}(\rho_{m})\mathcal{P}^{-1} & = & \rho_{m};\nonumber \\
\mathcal{P}(\overrightarrow{j_{e}})\mathcal{P}^{-1} & = & -\overrightarrow{j_{e}};\nonumber \\
\mathcal{P}(\phi)\mathcal{P}^{-1} & = & -\phi_{m};\nonumber \\
\mathcal{P}(\overrightarrow{\nabla})\mathcal{P}^{-1} & = & -\overrightarrow{\nabla};\nonumber \\
\mathcal{P}(\overrightarrow{C})\mathcal{P}^{-1} & = & -\overrightarrow{C};\nonumber \\
\mathcal{P}(\overrightarrow{E})\mathcal{P}^{-1} & = & -\overrightarrow{E};\label{eq:21}\end{eqnarray}
showing that due to change of sign, these quantity have odd parity.
Now applying the parity transformations to equation (\ref{eq:1}),
we find that the generalized Dirac - Maxwell's equations (\ref{eq:1})
are invariant under parity transformations. On applying the parity
transformation to equations (\ref{eq:4}, \ref{eq:6}, \ref{eq:7},
\ref{eq:9}, \ref{eq:10}), we get 

\begin{eqnarray}
\mathcal{P}(\overrightarrow{\psi})\mathcal{P}^{-1} & = & -\overrightarrow{\psi};\nonumber \\
\mathcal{P}(\phi)\mathcal{P}^{-1} & = & \phi;\nonumber \\
\mathcal{P}(\overrightarrow{V})\mathcal{P}^{-1} & = & -\overrightarrow{V};\nonumber \\
\mathcal{P}(\rho)\mathcal{P}^{-1} & = & \rho;\nonumber \\
\mathcal{P}(\overrightarrow{J})\mathcal{P}^{-1} & = & -\overrightarrow{J};\nonumber \\
\mathcal{P}(\overrightarrow{S})\mathcal{P}^{-1} & = & -\overrightarrow{S}.\label{eq:22}\end{eqnarray}
As such, the components of generalized complex potential given by
equations (\ref{eq:12}) are invariant under parity transformations.
The tensorial form of generalized Maxwell Dirac equation of dyons
in isotropic homogeneous medium given by equations (\ref{eq:13})
are also invariant under parity transformations. On the other hand,
the generalized field equations (\ref{eq:16}) and the Lorentz four
- force equation of motion (\ref{eq:18}) for generalized fields of
dyon in homogeneous isotropic medium loose their invariance under
parity transformations.

\section{Effect of Time reversal on Generalized fields of dyons}

In simple classical terms, time reversal just means replacing $t$
by $-t$, inverting the direction of the flow of time \cite{key-2}.
Reversing time also reverses the time derivatives of spatial quantities.
So, it reverses momentum and angular momentum. Thus the following
physical quantities like displacement ($x$), Acceleration ($a$),
electric charge source density ($\rho_{e}$), magnetic current source
density ($\overrightarrow{j_{m}}$), electric scalar potential ($\phi_{e}$),
magnetic vector potential ($\overrightarrow{D}$), space derivative
($\overrightarrow{\nabla}$) and generalized electric field ($\overrightarrow{E}$)
are unaffected and do not change \cite{key-16} their sign under time
reversal i.e.\begin{eqnarray}
\mathcal{T}(x)\mathcal{T}^{-1} & = & x;\nonumber \\
\mathcal{T}(\overrightarrow{a})\mathcal{T}^{-1} & = & \overrightarrow{a};\nonumber \\
\mathcal{T}(\rho_{e})\mathcal{T}^{-1} & = & \rho_{e};\nonumber \\
\mathcal{T}(\overrightarrow{j_{m}})\mathcal{T}^{-1} & = & \overrightarrow{j_{m}};\nonumber \\
\mathcal{T}(\phi_{e})\mathcal{T}^{-1} & = & \phi_{e};\nonumber \\
\mathcal{T}(\overrightarrow{D})\mathcal{T}^{-1} & = & \overrightarrow{D};\nonumber \\
\mathcal{T}(\overrightarrow{\nabla})\mathcal{T}^{-1} & = & \overrightarrow{\nabla};\nonumber \\
\mathcal{T}(\overrightarrow{E})\mathcal{T}^{-1} & = & \overrightarrow{E}.\label{eq:23}\end{eqnarray}
On the other hand the physical quantities like time ($t$), particle
velocity ($\overrightarrow{u}$),electric current source density ($\overrightarrow{j_{e}}$),
magnetic scalar potential ($\phi_{m}$), electric vector potential
($\overrightarrow{C}$), time derivative ($\partial_{t}$) and generalized
magnetic field ($\overrightarrow{B}$) are affected and change their
sign under time reversal i.e..

\begin{eqnarray}
\mathcal{T}(t)\mathcal{T}^{-1} & = & -t;\nonumber \\
\mathcal{T}(\overrightarrow{u})\mathcal{T}^{-1} & = & -\overrightarrow{u};\nonumber \\
\mathcal{T}(\overrightarrow{j_{e}})\mathcal{T}^{-1} & = & -\overrightarrow{j_{e}};\nonumber \\
\mathcal{T}(\phi_{m})\mathcal{T}^{-1} & = & -\phi_{m};\nonumber \\
\mathcal{T}(\overrightarrow{C})\mathcal{T}^{-1} & = & -\overrightarrow{C};\nonumber \\
\mathcal{T}(\partial_{t})\mathcal{T}^{-1} & = & -\partial_{t};\nonumber \\
\mathcal{T}(\overrightarrow{B})\mathcal{T}^{-1} & = & -\overrightarrow{B}.\label{eq:24}\end{eqnarray}
Applying the time reversal ($\mathcal{T}$) symmetry to equation (\ref{eq:1}),
we find that the generalized Dirac - Maxwell equations (\ref{eq:1})
in the homogeneous medium are invariant under $\mathcal{T}$ symmetry.
Similarly, on applying $\mathcal{T}$ symmetry to equations (\ref{eq:4},\ref{eq:6},\ref{eq:7},
\ref{eq:8},\ref{eq:10}), we get

\begin{eqnarray}
\mathcal{T}(\overrightarrow{\psi})\mathcal{T}^{-1} & =\overrightarrow{\psi} & ;\nonumber \\
\mathcal{T}(\phi)\mathcal{T}^{-1} & = & \phi;\nonumber \\
\mathcal{T}(\overrightarrow{V})\mathcal{T}^{-1} & = & -\overrightarrow{V};\nonumber \\
\mathcal{T}(\rho)\mathcal{T}^{-1} & = & \rho;\nonumber \\
\mathcal{T}(\overrightarrow{J})\mathcal{T}^{-1} & = & -\overrightarrow{J};\nonumber \\
\mathcal{T}(\overrightarrow{S})\mathcal{T}^{-1} & =\overrightarrow{S} & .\label{eq:25}\end{eqnarray}
Thus, the components of generalized complex potential defined by equation
(\ref{eq:12}) are invariant under $\mathcal{T}$ symmetry. As such,
the tensorial forms of generalized Dirac- Maxwell equation of dyons
in isotropic homogeneous medium given by equation (\ref{eq:13}) remain
invariant under the time reversal $(\mathcal{T})$ symmetry while
the generalized equations (\ref{eq:16}) and (\ref{eq:18}) describe
respectively the field equation and Lorentz force equation of motion
for dyons are not invariant under $\mathcal{T}$ symmetry like parity.

\section{Dyon fields under Charge Conjugation }

Classically, charge conjugation (Charge symmetry) replaces positive
charges by negative charges and vice versa \cite{key-3}. Since electric
and magnetic fields have their origins in charges, one can reverse
these fields on the application of charge conjugation. Charge conjugation
also involves reversing all the internal quantum numbers like lepton
number, baryon number and strangeness etc. The physical quantities
like displacement ($x$), time ($t$), mass ($m$), particle velocity
($u$), acceleration ($a$), space derivative ($\overrightarrow{\nabla}$)
and time derivative ($\partial_{t}$) are invariant under charge conjugation
\cite{key-16} while the energy, momentum or spin are unaffected i.e.

\begin{eqnarray}
\mathcal{C}(x)\mathcal{C}^{-1} & = & x;\nonumber \\
\mathcal{C}(t)\mathcal{C}^{-1} & = & t;\nonumber \\
\mathcal{C}(m) & \mathcal{C}^{-1}= & m;\nonumber \\
\mathcal{C}(\overrightarrow{u})\mathcal{C}^{-1} & = & \overrightarrow{u};\nonumber \\
\mathcal{C}(\overrightarrow{a})\mathcal{C}^{-1} & = & \overrightarrow{a};\nonumber \\
\mathcal{C}(\overrightarrow{\nabla})\mathcal{C}^{-1} & = & \overrightarrow{\nabla};\nonumber \\
\mathcal{C}(\partial_{t})\mathcal{C}^{-1} & = & \partial_{t}.\label{eq:26}\end{eqnarray}
 On the other hand the physical quantities like electric charge ($e$),
magnetic charge ($g$), electric charge source density ($\rho_{e}$),
magnetic charge source density ($\rho_{m}$), generalized charge source
density ($\rho$), electric scalar potential $(\phi_{e})$, magnetic
scalar potential $(\phi_{m})$, generalized scalar potential $(\phi)$,electric
current source density ($\overrightarrow{j_{e}}$), magnetic current
source density ($\overrightarrow{j_{m}}$), generalized current source
density ($\overrightarrow{j}$), electric vector potential ($\overrightarrow{C}$),
magnetic vector potential $(\overrightarrow{D})$, generalized vector
potential $(\overrightarrow{V})$, generalized electric field ($\overrightarrow{E}$)
and generalized magnetic field ($\overrightarrow{B}$) are directly
related to the charge and thus their sign is changed under charge
symmetry as,\begin{eqnarray}
\mathcal{C}(e)\mathcal{C}^{-1} & = & -e;\nonumber \\
\mathcal{C}(g)\mathcal{C}^{-1} & = & -g;\nonumber \\
\mathcal{C}(\rho_{e})\mathcal{C}^{-1} & = & -\rho_{e};\nonumber \\
\mathcal{C}(\rho_{m})\mathcal{C}^{-1} & = & -\rho_{m};\nonumber \\
\mathcal{C}(\rho)\mathcal{C}^{-1} & = & -\rho;\nonumber \\
\mathcal{C}(\phi_{e})\mathcal{C}^{-1} & = & -\phi_{e};\nonumber \\
\mathcal{C}(\phi_{m})\mathcal{C}^{-1} & = & -\phi_{m};\nonumber \\
\mathcal{C}(\phi)\mathcal{C}^{-1} & = & -\phi;\nonumber \\
\mathcal{C}(\overrightarrow{j_{e}})\mathcal{C}^{-1} & = & -\overrightarrow{j_{e}};\nonumber \\
\mathcal{C}(\overrightarrow{j_{m}})\mathcal{C}^{-1} & = & -\overrightarrow{j_{m}};\nonumber \\
\mathcal{C}(\overrightarrow{J})\mathcal{C}^{-1} & = & -\overrightarrow{J};\nonumber \\
\mathcal{C}(\overrightarrow{C})\mathcal{C}^{-1} & = & -\overrightarrow{C};\nonumber \\
\mathcal{C}(\overrightarrow{D})\mathcal{C}^{-1} & = & -\overrightarrow{D};\nonumber \\
\mathcal{C}(\overrightarrow{V})\mathcal{C}^{-1} & = & -\overrightarrow{V};\nonumber \\
\mathcal{C}(\overrightarrow{E})\mathcal{C}^{-1} & = & -\overrightarrow{E};\nonumber \\
\mathcal{C}(\overrightarrow{B})\mathcal{C}^{-1} & = & -\overrightarrow{B}.\label{eq:27}\end{eqnarray}
If we check the $\mathcal{C}$ invariance of equation (\ref{eq:1}),
we find that the generalized Dirac - Maxwell's equations (\ref{eq:1})
of dyons in the homogeneous medium are invariant under charge symmetry.
Similarly if we apply the $\mathcal{C}$ invariance conditions to
equations (\ref{eq:4}) and (\ref{eq:10}), we see that the physical
quantities like generalized field $(\overrightarrow{\psi})$ and generalized
field current $(\overrightarrow{S})$ of dyon in homogeneous medium
change their sign as, 

\begin{eqnarray}
\mathcal{C}(\overrightarrow{\psi})\mathcal{C}^{-1} & =-\overrightarrow{\psi} & ;\nonumber \\
\mathcal{C}(\overrightarrow{S})\mathcal{C}^{-1} & =-\overrightarrow{S} & .\label{eq:28}\end{eqnarray}
Applying the $\mathcal{C}$ transformation to, the complex potential
equations (\ref{eq:12}), the tensorial form of Dirac - Maxwell equations
(\ref{eq:13}), generalized field equation (\ref{eq:16}) and Lorentz
force equation (\ref{eq:18}) of dyons in isotropic homogeneous medium,
we see that these equations of dyons in homogeneous isotropic medium
are invariant under $\mathcal{C}$ transformations.

\section{Combined operation of Parity and Time Reversal $(\mathcal{PT})$
and dyon fields}

The most widely known symmetry are based on group theory is called
$\mathcal{PT}$ symmetry. In case of $\mathcal{PT}$ symmetry, the
particle velocity, potential and current are invariant under the simultaneous
action of the space and time reflection operators $\mathcal{P}$ and
$\mathcal{T}$ \cite{key-6} i.e. 

\begin{eqnarray}
(\mathcal{PT})(\overrightarrow{u})(\mathcal{P\mathcal{T}})^{-1} & = & \overrightarrow{u};\nonumber \\
(\mathcal{PT})(\rho_{e})(\mathcal{P\mathcal{T}})^{-1} & = & \rho_{e};\nonumber \\
(\mathcal{PT})(\rho_{m})(\mathcal{P\mathcal{T}})^{-1} & = & \rho_{m};\nonumber \\
(\mathcal{PT})(\rho)(\mathcal{P\mathcal{T}})^{-1} & = & \rho;\nonumber \\
(\mathcal{PT})(\overrightarrow{j_{e}})(\mathcal{P\mathcal{T}})^{-1} & = & \overrightarrow{j_{e}};\nonumber \\
(\mathcal{PT})(\overrightarrow{j_{m}})(\mathcal{P\mathcal{T}})^{-1} & = & \overrightarrow{j_{m}};\nonumber \\
(\mathcal{PT})(\overrightarrow{J})(\mathcal{P\mathcal{T}})^{-1} & = & \overrightarrow{J};\nonumber \\
(\mathcal{PT})(\phi_{e})(\mathcal{P\mathcal{T}})^{-1} & = & \phi_{e};\nonumber \\
(\mathcal{PT})(\phi_{m})(\mathcal{P\mathcal{T}})^{-1} & = & \phi_{m};\nonumber \\
(\mathcal{PT})(\phi)(\mathcal{P\mathcal{T}})^{-1} & = & \phi;\nonumber \\
(\mathcal{PT})(\overrightarrow{C})(\mathcal{P\mathcal{T}})^{-1} & = & \overrightarrow{C};\nonumber \\
(\mathcal{PT})(\overrightarrow{D})(\mathcal{P\mathcal{T}})^{-1} & = & \overrightarrow{D};\nonumber \\
(\mathcal{PT})(\overrightarrow{V})(\mathcal{P\mathcal{T}})^{-1} & = & \overrightarrow{V};\label{eq:29}\end{eqnarray}
where the symbols of physical quantities have their usual meaning
as discussed above. On the other hand, the following physical variables
with  their usual symbols \begin{eqnarray}
(\mathcal{PT})(x)(\mathcal{P\mathcal{T}})^{-1} & = & -x;\nonumber \\
(\mathcal{PT})(t)(\mathcal{P\mathcal{T}})^{-1} & = & -t;\nonumber \\
(\mathcal{PT})(\overrightarrow{a})(\mathcal{P\mathcal{T}})^{-1} & = & -\overrightarrow{a};\nonumber \\
(\mathcal{PT})(\overrightarrow{\nabla})(\mathcal{P\mathcal{T}})^{-1} & = & -\overrightarrow{\nabla};\nonumber \\
(\mathcal{PT})(\overrightarrow{E})(\mathcal{P\mathcal{T}})^{-1} & = & -\overrightarrow{E};\nonumber \\
(\mathcal{PT})(\overrightarrow{B})(\mathcal{P\mathcal{T}})^{-1} & = & -\overrightarrow{B};\label{eq:30}\end{eqnarray}
get changed their sign under $\mathcal{PT}$ symmetry. Simultaneously,
if we perform the $\mathcal{PT}$ symmetry operation to the equation
(\ref{eq:1}), we find that the generalized Dirac-Maxwell's equations
of dyon in the homogeneous medium are invariant under $\mathcal{PT}$
symmetry. Applying the $\mathcal{PT}$ symmetry to generalized field
$(\overrightarrow{\psi})$ and generalized field current $(\overrightarrow{S})$
of dyons in homogeneous medium respectively given by equations (\ref{eq:4})
and (\ref{eq:10}), we see that these quantities change their sign
under $\mathcal{PT}$ symmetry i.e.

\begin{eqnarray}
(\mathcal{PT})(\overrightarrow{\psi})(\mathcal{P\mathcal{T}})^{-1} & = & -\overrightarrow{\psi};\nonumber \\
(\mathcal{PT})(\overrightarrow{S})(\mathcal{P\mathcal{T}})^{-1} & = & -\overrightarrow{S};\label{eq:31}\end{eqnarray}
while the dyon field equations (\ref{eq:1}, \ref{eq:8}, and \ref{eq:12})
are invariant under $\mathcal{PT}$ symmetry. The combined operation
of $\mathcal{PT}$ transformation also shows the invariance of the
tensorial forms of Maxwell's equations (\ref{eq:13}) and the generalized
field equation (\ref{eq:16}) of dyons in homogeneous isotropic medium.
On the other hand, the Lorentz four - force equation (\ref{eq:18})
of motion for dyon in homogeneous isotropic medium looses its invariance
under $\mathcal{PT}$ symmetry and thus gets changed its sign under
this symmetry.

\section{Charge Conjugation and Parity for field associated with dyons}

Associated with the conservation laws which govern the behavior of
physical particles, charge conjugation $(\mathcal{C})$ and parity
$(\mathcal{P})$ combine to constitute a fundamental symmetry called
$\mathcal{CP}$ invariance \cite{key-3,key-4,key-24}. The following
quantities are invariant under the combined effect of $\mathcal{CP}$
i.e.

\begin{eqnarray}
(\mathcal{CP})(t)(\mathcal{CP})^{-1} & = & -t;\nonumber \\
(\mathcal{CP})(m)(\mathcal{CP})^{-1} & = & m;\nonumber \\
(\mathcal{CP})(\partial_{t})(\mathcal{CP})^{-1} & = & \partial_{t};\nonumber \\
(\mathcal{CP})(\rho_{m})(\mathcal{CP})^{-1} & = & \rho_{m};\nonumber \\
(\mathcal{CP})(\overrightarrow{j_{e}})(\mathcal{CP})^{-1} & = & \overrightarrow{j_{e}};\nonumber \\
(\mathcal{CP})(\phi_{m})(\mathcal{CP})^{-1} & = & \phi_{m};\nonumber \\
(\mathcal{CP})(\overrightarrow{C})(\mathcal{CP})^{-1} & = & \overrightarrow{C};\nonumber \\
(\mathcal{CP})(\overrightarrow{E})(\mathcal{CP})^{-1} & = & \overrightarrow{E}.\label{eq:32}\end{eqnarray}
On the other hand, we see that the following parameters change their
sign under the combined operation $\mathcal{CP}$ of charge conjugation
and parity i.e. 

\begin{eqnarray}
(\mathcal{CP})(x)(\mathcal{CP})^{-1} & = & -x;\nonumber \\
(\mathcal{CP})(\overrightarrow{u})(\mathcal{CP})^{-1} & = & -\overrightarrow{u};\nonumber \\
(\mathcal{CP})(e)(\mathcal{CP})^{-1} & = & -e;\nonumber \\
(\mathcal{CP})(g)(\mathcal{CP})^{-1} & = & -g;\nonumber \\
(\mathcal{CP})(\overrightarrow{\nabla})(\mathcal{CP})^{-1} & = & -\overrightarrow{\nabla};\nonumber \\
(\mathcal{CP})(\rho_{e})(\mathcal{CP})^{-1} & = & -\rho_{e};\nonumber \\
(\mathcal{CP})(\overrightarrow{j_{m}})(\mathcal{CP})^{-1} & = & -\overrightarrow{j_{m}};\nonumber \\
(\mathcal{CP})(\phi_{e})(\mathcal{CP})^{-1} & = & -\phi_{e};\nonumber \\
(\mathcal{CP})(\overrightarrow{D})(\mathcal{CP})^{-1} & = & -\overrightarrow{D};\nonumber \\
(\mathcal{CP})(\overrightarrow{B})(\mathcal{CP})^{-1} & = & -\overrightarrow{B}.\label{eq:33}\end{eqnarray}
Applying symmetry operation $\mathcal{CP}$ to equation (\ref{eq:1}),
we see that the generalized Maxwell's equation in the homogeneous
medium (\ref{eq:1}) are invariant under $\mathcal{CP}$ symmetry.
Applying the $\mathcal{CP}$ symmetry operation to equations (\ref{eq:4},\ref{eq:7},\ref{eq:9},\ref{eq:10})
, the nature of physical variables like generalized field $(\overrightarrow{\psi}),$
generalized scalar potential of dyon $(\phi)$, generalized vector
potential of dyon $(\overrightarrow{V})$, generalized charge density
of dyon $(\rho)$, generalized current density of dyon $(\overrightarrow{J})$
and generalized field current $(\overrightarrow{S})$ in homogeneous
medium are changed as follows,

\begin{eqnarray}
(\mathcal{CP})(\overrightarrow{\psi})(\mathcal{CP})^{-1} & =\overrightarrow{\psi} & ;\nonumber \\
(\mathcal{CP})(\phi)(\mathcal{CP})^{-1} & = & -\phi;\nonumber \\
(\mathcal{CP})(\overrightarrow{V})(\mathcal{CP})^{-1} & = & \overrightarrow{V};\nonumber \\
(\mathcal{CP})(\rho)(\mathcal{CP})^{-1} & = & -\rho;\nonumber \\
(\mathcal{CP})(\overrightarrow{J})(\mathcal{CP})^{-1} & = & \overrightarrow{J};\nonumber \\
(\mathcal{CP})(\overrightarrow{S})(\mathcal{CP})^{-1} & = & \overrightarrow{S}.\label{eq:34}\end{eqnarray}
As such, the complex potential field equations (\ref{eq:12}) and
the tensorial form of generalized Maxwell Dirac equation of dyons
(\ref{eq:13}) in isotropic homogeneous medium are invariant under
$\mathcal{CP}$ transformations. But the generalized field equation
(\ref{eq:16}) and the Lorentz four - force equation of motion (\ref{eq:18})
of dyons in homogeneous isotropic medium are not invariant under $\mathcal{CP}$
symmetry.

\section{CPT and fields associated with dyons}

We have seen that there are three symmetries which usually, but not
always, hold are those of charge conjugation $(\mathcal{C})$, parity
$(\mathcal{P})$ and time reversal $(\mathcal{T})$. Examples in nature
can be cited for the violation of each of these symmetries individually.
It was thought for a time that $\mathcal{CP}$ (parity transformation
plus charge conjugation) would always leave a system invariant, but
the notable example of the neutral kaons has shown a slight violation
of $\mathcal{CP}$ symmetry. We are left with the combination of all
three, i.e. $\mathcal{CPT}$, a profound symmetry which is assumed
to be consistent with all known experimental observations. The velocity
of the particle $(\overrightarrow{u})$, electric field $\overrightarrow{(E})$
and the magnetic field $(\overrightarrow{B})$ are invariant under
the combined operation of $\mathcal{CPT},$

\begin{eqnarray}
(\mathcal{CPT})(\overrightarrow{u})(\mathcal{CP}T)^{-1} & = & \overrightarrow{u};\nonumber \\
(\mathcal{CPT})(\overrightarrow{E})(\mathcal{CP}T)^{-1} & = & \overrightarrow{E};\nonumber \\
(\mathcal{CPT})(\overrightarrow{B})(\mathcal{CP}T)^{-1} & = & \overrightarrow{B}.\label{eq:35}\end{eqnarray}
On the other hand, other physical variables are not invariant under
the combined operation of $\mathcal{CPT}$ , i.e.

\begin{eqnarray}
(\mathcal{CPT})(x)(\mathcal{CP}T)^{-1} & = & -x;\nonumber \\
(\mathcal{CPT})(t)(\mathcal{CP}T)^{-1} & = & -t;\nonumber \\
(\mathcal{CPT})(e)(\mathcal{CP}T)^{-1} & = & -e;\nonumber \\
(\mathcal{CPT})(g)(\mathcal{CP}T)^{-1} & = & -g;\nonumber \\
(\mathcal{CPT})(\overrightarrow{\nabla})(\mathcal{CP}T)^{-1} & = & -\overrightarrow{\nabla};\nonumber \\
(\mathcal{CPT})(\partial_{t})(\mathcal{CP}T)^{-1} & = & -\partial_{t};\nonumber \\
(\mathcal{CPT})(\rho_{e})(\mathcal{CP}T)^{-1} & = & -\rho_{e};\nonumber \\
(\mathcal{CPT})(\rho_{m})(\mathcal{CP}T)^{-1} & = & -\rho_{m};\nonumber \\
(\mathcal{CPT})(\overrightarrow{j_{e}})(\mathcal{CP}T)^{-1} & = & -\overrightarrow{j_{e}};\nonumber \\
(\mathcal{CPT})(\overrightarrow{j_{m}})(\mathcal{CP}T)^{-1} & = & -\overrightarrow{j_{m}};\nonumber \\
(\mathcal{CPT})(\phi_{e})(\mathcal{CP}T)^{-1} & = & -\phi_{e};\nonumber \\
(\mathcal{CPT})(\phi_{m})(\mathcal{CP}T)^{-1} & = & -\phi_{m};\nonumber \\
(\mathcal{CPT})(\overrightarrow{C})(\mathcal{CP}T)^{-1} & = & -\overrightarrow{C};\nonumber \\
(\mathcal{CPT})(\overrightarrow{D})(\mathcal{CP}T)^{-1} & = & -\overrightarrow{D}.\label{eq:36}\end{eqnarray}
Consequently, if we apply the $\mathcal{CPT}$ operation to equations
(\ref{eq:4},\ref{eq:7},\ref{eq:9},\ref{eq:10}), we observe that
the quantities like generalized field $(\overrightarrow{\psi})$ and
generalized field current $(\overrightarrow{S})$ of dyons in homogeneous
medium are invariant while generalized scalar potential of dyon $(\phi)$,
generalized vector potential $(\overrightarrow{V})$ , generalized
charge density $(\rho)$ and generalized current density $(\overrightarrow{J})$
of dyon in homogeneous medium change are no more invariant i.e. 

\begin{eqnarray}
(\mathcal{CPT})(\overrightarrow{\psi})(\mathcal{CP}T)^{-1} & =\overrightarrow{\psi} & ;\nonumber \\
(\mathcal{CPT})(\phi)(\mathcal{CP}T)^{-1} & = & -\phi;\nonumber \\
(\mathcal{CPT})(\overrightarrow{V})(\mathcal{CP}T)^{-1} & = & -\overrightarrow{V};\nonumber \\
(\mathcal{CPT})(\rho)(\mathcal{CP}T)^{-1} & = & -\rho;\nonumber \\
(\mathcal{CPT})(\overrightarrow{J})(\mathcal{CP}T)^{-1} & = & -\overrightarrow{J};\nonumber \\
(\mathcal{CPT})(\overrightarrow{S})(\mathcal{CP}T)^{-1} & =\overrightarrow{S} & .\label{eq:37}\end{eqnarray}
Accordingly, the complex potential field equations (\ref{eq:12}),
the tensorial form of generalized Maxwell Dirac equations (\ref{eq:13},
\ref{eq:16}) and the Lorentz four - force equation of motion (\ref{eq:23})
of dyon in homogeneous isotropic medium are invariant under the combined
operation $(\mathcal{CPT})$.

\section{Conclusion}

From the foregoing analysis, we have observed that the Maxwell's -
Dirac equation (\ref{eq:1}) are invariant under discrete symmetries
like parity $\mathcal{P}$, time reversal $\mathcal{T}$, charge conjugation
$(\mathcal{C})$ , parity - time reversal operation $\mathcal{PT}$
, charge conjugation - parity operation $\mathcal{CP}$ and combined
operation $(\mathcal{CPT})$. We have also seen that the complex nature
of generalized field equation (\ref{eq:16}) and the Lorentz force
equation of motion (\ref{eq:18}) of dyon in homogeneous medium loose
their invariance under the discrete symmetries like parity $\mathcal{P}$,
Time reversal $\mathcal{T}$, parity - time reversal operation $\mathcal{PT}$
and charge conjugation - parity operation $\mathcal{CP}$. On the
other hand, these equations are invariant under the combined operation
$(\mathcal{CPT})$. In other words, we may conclude that $\mathcal{CPT}$
invariance is an exact symmetry for generalized fields of dyons and
reproduces the consistent theory of electric charge (magnetic monopole)
in the absence magnetic monopole (electric charge ) on dyons. As such
the statement of Ludens - Pauli - Schwinger \cite{key-25} that invariance
under Lorentz transformations implies $\mathcal{CPT}$ invariance
holds good for the case of dyons in homogeneous isotropic medium. 

\textbf{\underbar{ACKNOWLEDGMENT}}\textbf{:- }One of PSB is thankful
to Chinese Academy of Sciences and Third World Academy of Sciences
for awarding him CAS - TWAS Visiting Scholar Fellowship to pursue
a research program in China.

\end{document}